%% file: paper.tex
\documentclass[runningheads]{llncs}
\usepackage[T1]{fontenc}
\usepackage{graphicx}

\usepackage{makecell}
\usepackage{multirow}
\usepackage{amsmath}
\usepackage{mathrsfs}
\usepackage[bookmarks=true]{hyperref}
\usepackage{lipsum}
\usepackage{xcolor}
\usepackage{bbding}

\definecolor{gray}{RGB}{128,128,128}
\SetLipsumParListSurrounders{\begingroup\color{gray}}{\endgroup}
\begin{document}
\title{Label Propagation for 3D Carotid Vessel Wall Segmentation and Atherosclerosis Diagnosis}
\titlerunning{Carotid Vessel Wall Segmentation and Atherosclerosis Diagnosis}
%
\author{
Shishuai Hu
\and
Zehui Liao
\and
Yong Xia\Envelope
} 
\authorrunning{S. Hu et al.}
\institute{National Engineering Laboratory for Integrated Aero-Space-Ground-Ocean Big Data Application Technology, School of Computer Science and Engineering, Northwestern Polytechnical University, Xi’an 710072, China \\
\email{yxia@nwpu.edu.cn}}
\maketitle              
%

\input{main.tex}

\end{document}

%% file: main.tex
\begin{abstract}
Carotid vessel wall segmentation is a crucial yet challenging task in the computer-aided diagnosis of atherosclerosis.
Although numerous deep learning models have achieved remarkable success in many medical image segmentation tasks, accurate segmentation of carotid vessel wall on magnetic resonance (MR) images remains challenging, due to limited annotations and heterogeneous arteries.
In this paper, we propose a semi-supervised label propagation framework to segment lumen, normal vessel walls, and atherosclerotic vessel wall on 3D MR images.
By interpolating the provided annotations, we get 3D continuous labels for training 3D segmentation model.
With the trained model, we generate pseudo labels for unlabeled slices to incorporate them for model training.
Then we use the whole MR scans and the propagated labels to re-train the segmentation model and improve its robustness.
We evaluated the label propagation framework on the CarOtid vessel wall SegMentation and atherosclerOsis diagnosiS (COSMOS) Challenge dataset and achieved a QuanM score of 83.41\% on the testing dataset, which got the 1-st place on the online evaluation leaderboard. 
The results demonstrate the effectiveness of the proposed framework.

\end{abstract}
\section{Introduction}
\vspace{-0.2cm}
Atherosclerosis is a leading cause of death worldwide, which occurs with luminal narrowing and plaque formation in multiple vascular beds, including carotid arteries. 
Carotid vessel wall segmentation using magnetic resonance (MR) black-blood vessel wall imaging (BB-VWI) provides valuable information on normal and diseased arteries and characterizes atherosclerotic lesions, playing an essential role in early detection and treatment of carotid atherosclerosis to prevent the progression of cardiovascular disease~\cite{grand_challenge_vesselwall}.
Since manual segmentation of vessel walls is time-consuming and requires high concentration and expertise, automated segmentation methods are highly demanded to accelerate this process.
However, this task remains challenging mainly due to two reasons: (1) only limited and discontinuous well annotated slices can be accessed, resulting in the degradation of model's robustness; and
(2) the appearance of healthy and diseased arteries is heterogeneous, leading to the decrease of the model's segmentation performance on carotid plaques.

To address these issues, we formulate the carotid vessel wall segmentation and atherosclerosis diagnosis task as a 3D segmentation task that segments lumen, normal vessel wall, and atherosclerotic vessel wall from 3D MR scans directly, since the 3D structures of lumen, normal vessel wall, and even atherosclerotic vessel wall should be well preserved.
However, the discontinuous annotation renders as an obstacle for 3D structures segmentation, since only small part of slices in the MR scan were annotated with contours and atherosclerotic status.
To solve this issue, we interpolate the annotations first to convert 2D discontinuous annotations to 3D rough vessels.
Even though, the interpolated annotations only cover about 50 slices above and below the carotid bifurcation respectively, whereas there are about 250 slices in one MR scan, which means more than half of the slices are still unlabeled.
To fully utilize the unlabeled data and sufficiently train the model to learn the 3D structures, we adopt the simple but effective semi-supervised self-training strategy to propagate the annotations to the whole MR scan, and utilize the propagated annotations to re-train the 3D segmentation model.
We have evaluated the proposed label propagation framework on the COSMOS Challenge dataset and achieved a QuanM score of 83.41\% on the testing dataset, which got the 1-st place on the online evaluation leaderboard.

\vspace{-0.2cm}
\section{Dataset}
\vspace{-0.2cm}
The partially annotated carotid vessel wall segmentation and atherosclerosis diagnosis dataset provided by COSMOS 2022 challenge\footnote{\url{https://vessel-wall-segmentation-2022.grand-challenge.org/}}~\cite{grand_challenge_vesselwall} was used for this study.
The COSMOS dataset contains 75 MR scans with lumen contours, vessel wall contours, and atherosclerotic vessel wall classification labels annotated on only about 20\% MR slices with enough quality for vessel wall diagnosis.
Among all the MR scans, 50 cases are provided as training/validation cases, and the left 20 cases are used for online evaluation.
Only the annotations of training/validation cases are publicly available, while the annotations of testing cases cannot be accessed.

\vspace{-0.2cm}
\section{Method}
\vspace{-0.2cm}
\begin{figure}[t]
  \centering
  \includegraphics[width=1\textwidth]{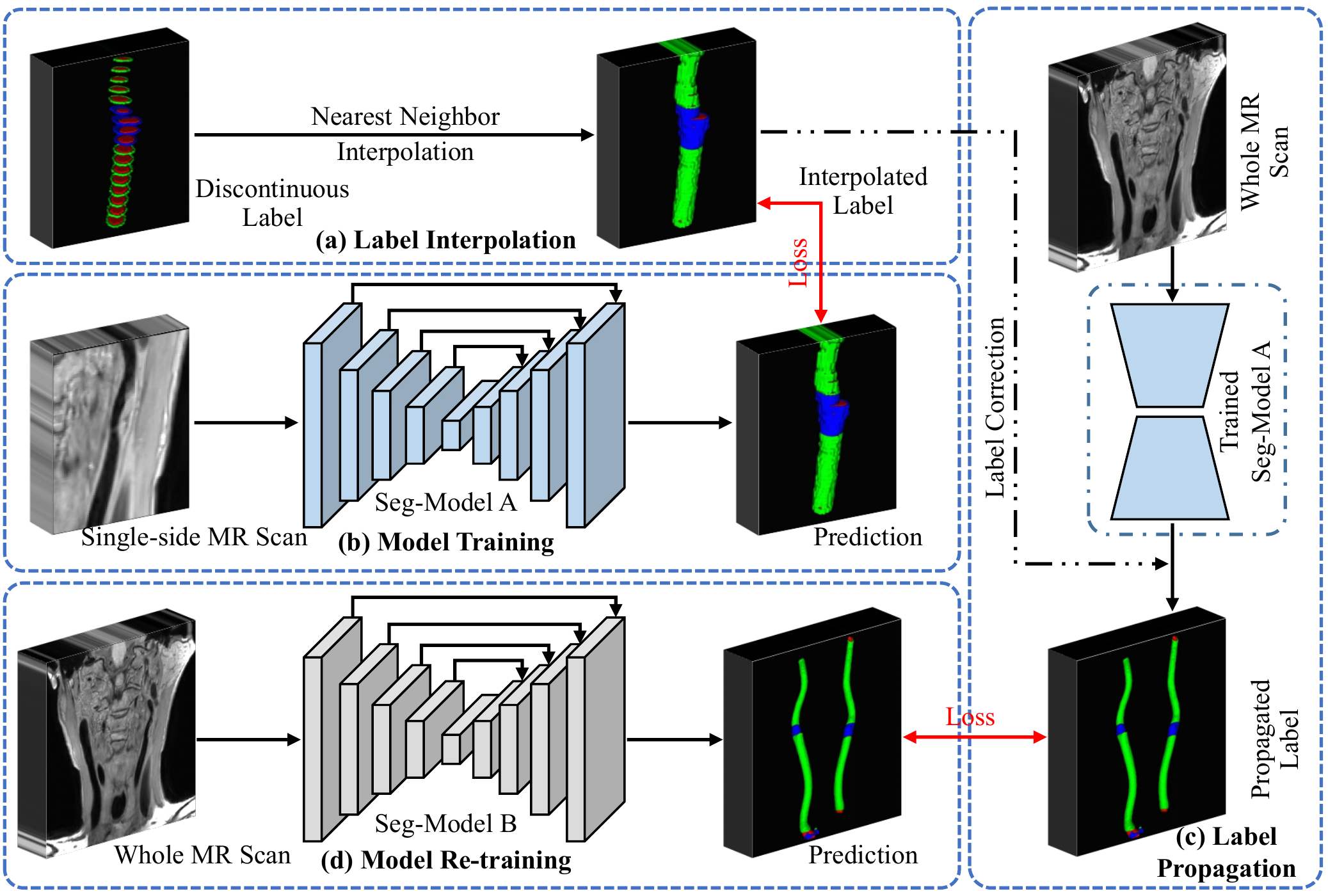}
  \caption{Illustration of label propagation for 3D carotid vessel wall segmentation and atherosclerosis diagnosis. Seg-Model-A and Seg-Model-B represent segmentation models. Only Seg-Model-B will be used for inference.}
  \label{fig:overview}
  \vspace{-0.6cm}
\end{figure}

The workflow of the proposed label propagation framework is mainly composed of (a) label interpolation, (b) model training, (c) label propagation, and (d) model re-training, as shown in Fig.~\ref{fig:overview}. We now delve each part into details.

\noindent\textbf{Label Definition and Interpolation.}
To relieve the issue caused by heterogeneity between the healthy and diseased vessel walls, we split the vessel walls into normal vessel walls ($i.e.$, labels in green color in Fig.~\ref{fig:overview}) and atherosclerotic vessel walls ($i.e.$, labels in purple color in Fig.~\ref{fig:overview}).
The lumen areas are represented as labels in red color in Fig.~\ref{fig:overview}.
Therefore, the segmentation model is request to segment lumen, normal vessel walls, and atherosclerotic vessel walls. 
As a result, the segmentation model can perform both carotid vessel wall segmentation and atherosclerosis diagnosis simultaneously.
We use Nearest Neighbor Interpolation to convert the 2D discontinuous labels to 3D vessels, as shown in Fig.~\ref{fig:overview} (a).

\noindent\textbf{Model Training.}
We use nnUNet~\cite{isensee_nnu-net_2021} as the backbone to perform 3D segmentation using the MR scans and corresponding interpolated labels.
Since the original left carotid vessel and right carotid vessel were annotated respectively, the annotated indexes of them might be different, leading to some missing annotations at the start or the end of the left or the right vessels.
Therefore, we split the whole MR scan into single-side MR scans to train the model Seg-Model-A.

\noindent\textbf{Label Propagation.}
With the help of the trained Seg-Model-A, we can generate pseudo labels for the whole MR scans.
However, the generated pseudo labels might not be accurate enough.
Therefore, we use the interpolated labels to correct the pseudo labels, as shown in Fig.~\ref{fig:overview} (c).

\noindent\textbf{Model Re-training.}
Using the propagated labels, we re-train a 3D nnUNet model ($i.e.$, Seg-Model-B in Fig.~\ref{fig:overview} (d)) with whole MR scans.
Since more data can be used for training, the robustness of Seg-Model-B can be further improved.

\noindent\textbf{Inference.}
Given a test image, we only use Seg-Model-B to perform inference in a sliding window manner.
Then we can get lumen, normal vessel walls, and atherosclerotic vessel walls, of which the latter two are regarded as vessel walls for the segmentation task.
For the atherosclerosis diagnosis task, we split the whole MR scan into single-side MR scans, and classified the single-side MR slice into normal or atherosclerosis by judging whether the MR slice contains atherosclerotic vessel walls.

\vspace{-0.2cm}
\section{Experiments and Results}
\vspace{-0.2cm}
\noindent\textbf{Implementation Details and Evaluation Metric}
We applied data augmentation techniques, including random cropping, random rotation, random scaling, random flipping, random Gaussian noise addition, and elastic deformation to increase the diversity of training data.
The MR scans were normalized by subtracting their mean and dividing by their standard deviation.
The size of the input scan volume  was set to $96\times160\times160$, and the batch size was set to 2.
Totally 5 down-samplings were performed in both Seg-Model-A and Seg-Model-B. 
The SGD algorithm was used as the optimizer.
The initial learning rate $lr_0$ was set to 0.01 and decayed according to $lr = lr_0\times(1-t/T)^{0.9}$, where $t$ is the current epoch and $T=500$ is the maximum epoch.
The combination of Dice loss and Cross-Entropy loss was used as loss functions.
The whole framework was implemented in PyTorch and trained using an NVIDIA 2080Ti.
We use Dice Similarity Coefficient (DSC) to evaluate the segmentation performance on the training set with 4-fold cross validation.

\vspace{-0.2cm}
\subsection{Results}
\vspace{-0.2cm}

\begin{table}[]
\centering
\caption{Performance (DSC \%) of Seg-Model-A, Seg-Model-B, and 2D-Seg. The best results are highlighted with \textbf{bold}.}
\label{tab:results}
\setlength\tabcolsep{5pt}
\renewcommand{\arraystretch}{1.2}
\begin{tabular}{l|c|c|c|c}
\hline \hline
Methods & Lumen & Normal Vessel Wall & Diseased Vessel Wall & Average \\
\hline
2D-Seg  & 87.87 & 63.42 & 46.39 & 65.89 \\
\hline
Seg-Model-A & 92.34 & 66.37 & 50.61 & 69.77 \\
\hline
Seg-Model-B & \textbf{93.47} & \textbf{87.23} & \textbf{72.71} & \textbf{84.47} \\
\hline \hline
\end{tabular}
\vspace{-0.6cm}
\end{table}

We compare the the segmentation performance of Seg-Model-A, Seg-Model-B, and the 2D segmentation model trained using the interpolated labels ($i.e.$, 2D-Seg), as shown in~\tablename{~\ref{tab:results}}.
It reveals that 3D segmentation models are superior to 2D models on vessel segmentation.
Also, \tablename{~\ref{tab:results}} shows that Seg-Model-B improves the segmentation accuracy a lot than Seg-Model-A, confirming the effectiveness of the label propagation strategy.

\vspace{-0.2cm}
\section{Conclusion}
\vspace{-0.2cm}
In this work, we propose a label propagation framework for carotid vessel wall segmentation and atherosclerosis diagnosis. By formulating this task as a 3D segmentation task, we can solve both segmentation and diagnosis in a unified 3D framework. 
Moreover, the label propagation strategy provides sufficient labeled 3D data for model training and hence improve model's robustness.
Experimental results on the COSMOS challenge dataset demonstrate the effectiveness of the proposed framework.
In our future work, we will improve the label propagation strategy to fully utilized the unlabeled MR slices.

\clearpage
\bibliographystyle{splncs04}
\bibliography{reference}

%% file: paper.bbl
\begin{thebibliography}{1}
\providecommand{\url}[1]{\texttt{#1}}
\providecommand{\urlprefix}{URL }
\providecommand{\doi}[1]{https://doi.org/#1}

\bibitem{grand_challenge_vesselwall}
Chen, H., Zhao, X., Dou, J., Du, C., Yang, R., Sun, H., Yu, S., Zhao, H., Yuan,
  C., Balu, N.: Carotid vessel wall segmentation and atherosclerosis diagnosis
  challenge (2022),
  \url{https://vessel-wall-segmentation-2022.grand-challenge.org/}

\bibitem{isensee_nnu-net_2021}
Isensee, F., Jaeger, P.F., Kohl, S.A.A., Petersen, J., Maier-Hein, K.H.:
  {nnU}-net: a self-configuring method for deep learning-based biomedical image
  segmentation  \textbf{18}(2),  203--211. \doi{10.1038/s41592-020-01008-z},
  \url{https://www.nature.com/articles/s41592-020-01008-z}

\end{thebibliography}
